\documentclass[%
reprint,
nofootinbib,
aps,
twocolumn,
prd
]{revtex4-2}

\usepackage{amsmath,amssymb,amsbsy,amstext,amsthm,simplewick,amsfonts,tensor}
\usepackage[normalem]{ulem}
\usepackage{graphicx} % Include figure files
\usepackage{dcolumn}% Align table columns on decimal point
\usepackage{bm}% bold math
\usepackage{comment}
\usepackage{hyperref}
\usepackage{bbold}
\usepackage{hyperref}

\newcommand{\bPsi}{{\bm{\mathsf{\Psi}}}}
\newcommand{\bepsilon}{{\bm{\mathsf{\epsilon}}}}
\newcommand{\bS}{{\bm{S}}}
\newcommand{\bW}{{\bm{W}}}

\newcommand{\Vnl}{V_{\rm{nl}}}

\hypersetup{
	colorlinks = true,
	linkcolor = blue,
	citecolor = blue,
	urlcolor = blue
}

\def\l{\left}
\def\r{\right}
\def\({\l(}
\def\){\r)}
\def\[{\l[}
\def\]{\r]}
\def\pd{\partial}

\def\b{\boldsymbol}

\def\cal{\mathcal}

\def\rm{\mathrm}

\def\bx{\b x}
\def\by{\b y}
\def\bz{\b z}

\begin{document}

\title{Polarized Vector Oscillons}% Force line breaks with \\

\author{Hong-Yi Zhang}
\email{hongyi@rice.edu}
\author{Mudit Jain}
\email{mudit.jain@rice.edu}
\author{Mustafa A. Amin}%
\email{mustafa.a.amin@rice.edu}
\affiliation{Department of Physics and Astronomy, Rice University, Houston, TX 77005, USA}%

%\date{\today}

\begin{abstract}
Oscillons are spatially localized, time-periodic and long-lived configurations that were primarily proposed in scalar field theories with attractive self-interactions. In this paper, we demonstrate that oscillons also exist in the low-energy effective theory of an interacting massive (real) vector field. We provide two types of vector oscillons with vanishing orbital angular momentum, and approximately spherically symmetric energy density, but not field configurations. These are: (1) ``directional'' oscillons (linearly polarized), with vanishing total intrinsic spin, and (2) ``spinning'' oscillons (circularly polarized) with a macroscopic instrinsic spin equal to $\hbar\times$  number of particles in the oscillon. In contrast to the case with only gravitational interactions, the two oscillons have different energy at a fixed particle number even in the nonrelativistic limit. By carrying out relativistic $3+1$d simulations, we show that these oscillons can be long-lived (compared to the oscillation time for the fields), and can arise from a range of Gaussian initial spatial profiles. These considerations make vector oscillons potentially relevant during the early universe and in dark photon dark matter, with novel phenomenology related to their polarization.
\end{abstract}

\maketitle

%\tableofcontents

\section{Introduction}
Non-topological solitons arise due to a balance between attractive self-interaction and dispersion in field theories. They have been realized in a broad variety of contexts in nature, from water waves \cite{Russel:old} to Bose-Einstein condensates (BEC)~\cite{2002Natur.417..150S,2006PhRvL..96q0401C,Nguyen:2014, Luo2020}. They might also play a role in astrophysics and cosmology, with novel signatures in phase transitions in the early universe \cite{Copeland:1995fq,Kusenko:1997hj,Gleiser:2008dt,Amin:2011hj,Bond:2015zfa}, in the formation of structure and gravitational clustering \cite{Amin:2011hu,Olle:2019kbo, Amin:2019ums,Arvanitaki:2019rax,Niemeyer:2019gab,Marsh:2015wka, Levkov:2018kau},  production of gravitational waves~\cite{Zhou:2013tsa, Palenzuela:2017kcg, Liu:2017hua, Helfer:2018vtq,Croon:2018ftb, Amin:2018xfe, Dietrich:2018jov, Lozanov:2019ylm,Hiramatsu:2020obh} and electromagnetic radiation \cite{Hertzberg:2020dbk,Amin:2020vja,Levkov:2020txo,Amin:2021tnq}, formation of black holes~\cite{Helfer:2016ljl, Cotner:2018vug, Muia:2019coe, Widdicombe:2019woy, Nazari:2020fmk}, and even play a role in baryogenesis~\cite{Enqvist:1997si,Lozanov:2014zfa}. While massive scalar fields and their solitons have been explored extensively in the literature (for reviews, see \cite{Lee:1991ax, Liebling:2012fv, Nugaev:2019vru}), Nature provides us with many examples of higher spin fields. For instance, $W$ and $Z$ bosons in the Standard Model of particle physics, or  speculatively, as (some or all of) dark matter \cite{Graham:2015rva, Co:2018lka, Bastero-Gil:2018uel, Agrawal:2018vin, Kolb:2021xfn, Kolb:2021nob, Babichev:2016bxi, Alexander:2020gmv, ParticleDataGroup:2020ssz}.

\begin{figure}[t!]
	\centering
	\includegraphics[width=1\linewidth]{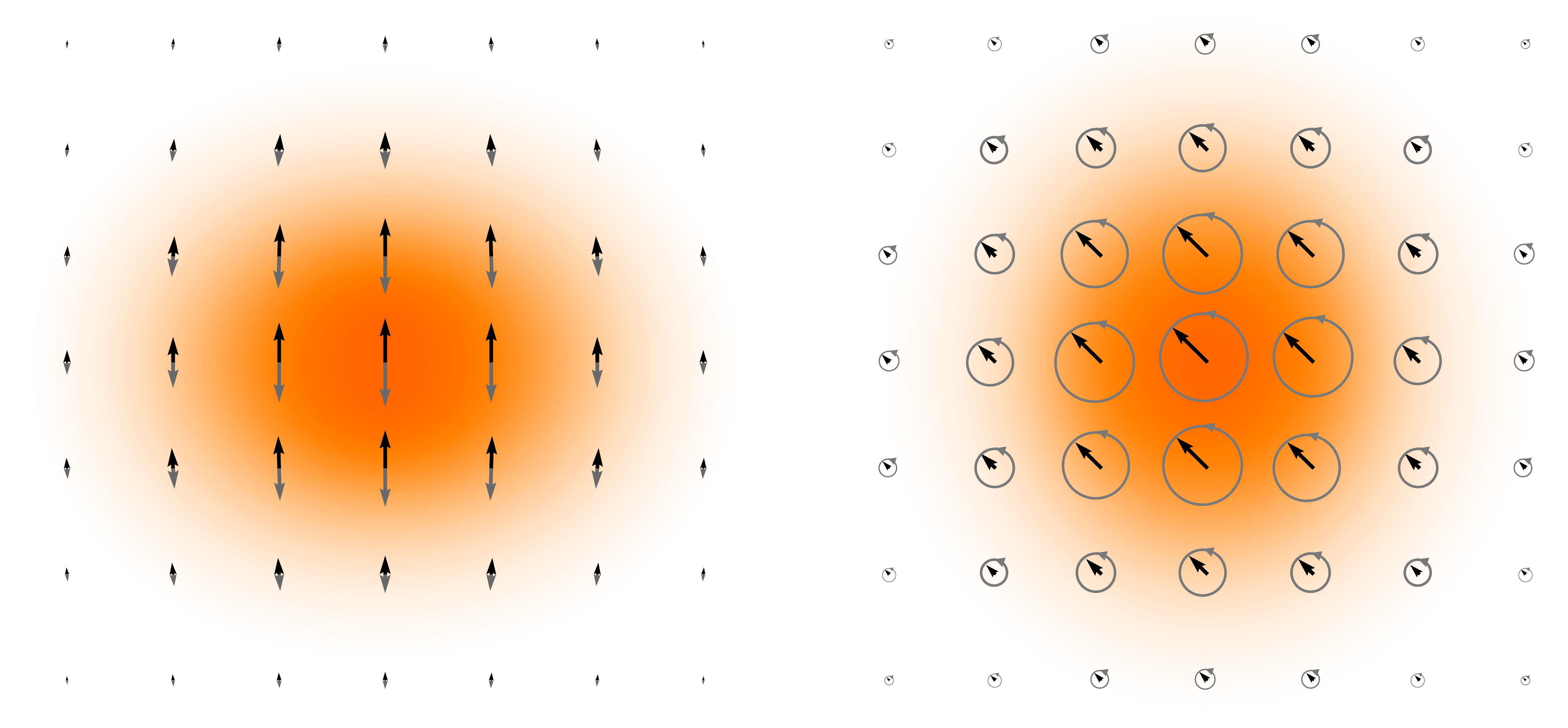}
	\caption{The directional and spinning oscillons obtained from relativistic simulations. The energy densities are approximately spherically symmetric, but the field configurations are not. For the spinning oscillon (right), the vector field at each point moves in a circle, resulting in a macroscopic intrinsic spin. For the directional oscillon (left), the field oscillates along an approximately fixed direction and has zero spin.}
	\label{fig:vectoroscillon}
\end{figure}

In this paper we study non-topological solitons in real-valued massive vector fields with attractive self-interactions. These spatially localized solitons are ``maximally'' polarized (with respect to a particular direction), i.e. either the vector field  configuration is primarily linearly polarized which we call a ``directional'' oscillon, or it is mostly circularly polarized that we refer to as a ``spinning'' oscillon (see Fig.~\ref{fig:vectoroscillon} for a quick description). Such objects might be present in the post-inflationary universe or constitute part of the present-day dark matter, and can provide novel gravitational and non-gravitational signatures revealing the intrinsic spin of the underlying massive (dark) vector field.
%~~~~~~~~~~~~~~~~~~~~~~~~~~
%~~~~~~~~~~~~~~~~~~~~~~~~~~

Although vector solitons can be supported solely by gravitational interactions \cite{Adshead:2021kvl,Jain:2021pnk}, self-interactions may appear naturally in the low-energy limit of an interacting vector field theory and play an important role in their phenomenology. For example, in the early universe, they can have a dominant effect in early structure formation~\cite{Gleiser:2010qt, Amin:2011hj, Sang:2020kpd, Padilla:2021zgm}. As we will show, self-interactions can also explicitly lift the degeneracy in energy between the directional and spinning oscillons, potentially determining which type of oscillon can form more easily. Furthermore, matter-wave solitons in BECs \cite{2002Natur.417..150S,2006PhRvL..96q0401C,Nguyen:2014, Luo2020} and electromagnetic solitons in nonlinear media (optical fibres)~\cite{menyuk1987nonlinear, christodoulides1988vector, rand2007observation, sun2009bound, baronio2012solutions} owe their existence to attractive self-interactions.

In what follows, we begin by studying vector oscillons using nonrelativistic approximations, and then perform fully relativistic numerical simulations to confirm their stability, longevity, etc. Finally, we summarize our results and discuss potential implications. Additional details and results are provided in the appendix. We work in natural units and adopt mostly plus signature for the metric.

\section{Model}
We study a real-valued massive spin-1 field $W_\mu$ with the Lagrangian
\begin{align}\label{eq:L_full}
\cal L = -\frac{1}{4} X_{\mu\nu}X^{\mu\nu} - V(W_\mu W^\mu)\,,
\end{align}
where $X_{\mu\nu} = \pd_\mu W_\nu - \pd_\nu W_\mu$ and the potential
\begin{align}
\label{potential}
V(W_\mu W^\mu) &= \frac{m^2}{2} W_\mu W^\mu - \frac{\lambda}{4} (W_\mu W^\mu)^2 + \frac{\gamma}{6} (W_\mu W^\mu)^3\nonumber\\
&\quad + ...
\end{align}
with positive couplings $\lambda$ and $\gamma$. Such effective potentials, for example, could arise in the low-energy regime of interacting massive vector fields \footnote{For definiteness we set $\gamma=\lambda^2/m^2$, and we have verified that qualitative features of vector oscillons remain unchanged if this value is changed by $\cal O(\lambda^2/m^2)$. In the appendix \ref{app:Effective_action}, we show that such an attractive potential indeed arises in the low-energy limit of a $U(1)$ Abelian-Higgs model with a heavy Higgs. For the sextic (and potentially higher) interaction term(s), however, we remain agnostic as to how they can be generated.}. The Euler-Lagrange equations are
\begin{align}
\label{EOM1}
&\nabla^2 W_0 - \pd_t \nabla\cdot \b W -2V'(W_\mu W^\mu) W_0 = 0,\\
\label{EOM2}
&\pd_t^2\b W - \pd_t \nabla W_0 + \nabla\times(\nabla\times \b W) + 2V'(W_\mu W^\mu) \b W = 0 .\nonumber
\end{align}

Using the Noether energy-momentum tensor  $T^{\mu\nu}= \partial^\nu W_\sigma X^{\mu\sigma}+ \eta^{\mu\nu}\cal L$, the energy $E \equiv \int\mathrm{d}^3x\,T^{00}$ is given by
\begin{equation}
\label{eq:E_relativistic}
E \!=\! \int\mathrm{d}^3x\,\Bigg[\frac{1}{2}\left(\dot{\b W} - \nabla W_0\right)^{\!2} + \frac{1}{2}\left(\nabla\times \b W\right)^{\!2}+2W^2_0V'+V\Bigg],
\end{equation}
where we have used the equations of motion and also discarded a boundary term to get the explicit expression above. Furthermore, the conserved 4-current associated with Lorentz invariance is $\cal M^{\mu\nu\sigma}=\cal L^{\mu\nu\sigma} + \cal S^{\mu\nu\sigma}$. We have separated out $\cal L^{\mu\nu\sigma}=x^\nu T^{\mu\sigma} - x^\sigma T^{\mu\nu}$ and $\cal S^{\mu\nu\sigma}= X^{\mu\nu}W^\sigma - X^{\mu\sigma}W^\nu$ so that the orbital and spin angular momentum densities are $L_i = (1/2) \epsilon_{ijk} \cal L^{0jk}$ and $S_i = (1/2) \epsilon_{ijk} \cal S^{0jk}$ respectively.  In particular, the spin density is
\begin{align}
\label{eq:spindensity}
    \b S = {\b W}\times\left(\dot{\b W} - \nabla W_0\right),
\end{align}
which will play a pivotal role in discriminating the directional and spinning oscillon configurations.

\section{Nonrelativistic limit}
It turns out to be sufficient to consider the nonrelativistic regime of the theory in the sense that  $|\nabla^2/m^2| \lesssim 10^{-2}$. We express the real vector field $\bW$ in terms of a complex vector field $\bPsi$, i.e.
\begin{align}
\label{nonrelativistic_expansion}
&\bW(t,\b x) \equiv \sqrt{\frac{2}{m}}\Re\left[\bPsi(t,\b x) e^{-imt}\right],
\end{align}
and $W_0(t,\bx)\equiv\sqrt{{2}/{m}}\,\Re\left[ \psi_0(t,\b x) e^{-imt}\right]$,
where the dependence of $\bPsi$ and $\psi_0$ on time is assumed to be weak. Upon plugging this expansion into the action, dropping all terms with the oscillatory factors $e^{\pm inmt}$ ($n \ge 2$), and keeping only the leading-order terms in time and spatial derivatives of $\bPsi$ (see, for example \cite{Namjoo:2017nia,Amin:2019ums,Salehian:2020bon,Salehian:2021khb,Adshead:2021kvl,Jain:2021pnk}), we get the following effective nonrelativistic Lagrangian density
\begin{align}\label{eq:NR_action}
    \cal L = \Re[i\bPsi^\dagger\dot{\bPsi}] - \frac{1}{2m}\nabla\bPsi^\dagger\cdot\nabla\bPsi - V_\mathrm{nl}(\bPsi^\dagger,\bPsi),
\end{align}
where we have solved for the constraint equation (to working order in $|\nabla^2/m^2|$), $\psi_0 = i\nabla\cdot\bPsi/m$ and the non-linear potential is
\begin{align}
\label{eq:Vnl}
    V_{\mathrm{nl}}(\bPsi^\dagger,\bPsi) &= -\frac{3\lambda}{8m^2}(\bPsi^\dagger\bPsi)^2 + \frac{5\gamma}{12m^3}(\bPsi^\dagger\bPsi)^3\nonumber\\
    &\quad + \left[\frac{\lambda}{8m^2}-\frac{\gamma}{4m^3}(\bPsi^\dagger\bPsi)\right](\bS\cdot\bS).
\end{align}
Note that we are able to write $V_{\rm {nl}}$ in terms of $\bPsi^\dagger\bPsi$, and the spin density  $\bm{S} = i\,\bPsi\times\bPsi^{\dagger}$,
which is the nonrelativistic, slowly varying part of \eqref{eq:spindensity}. This spin density can also be obtained directly from the rotational invariance of the nonrelativistic action for $\bPsi$. The appearance of $\bS\cdot\bS$ in $V_\rm{nl}$ suggests that the spin density will play a role in determining the energy of our solutions. This energy is given by
\begin{align}
\mathcal{E}=\int \mathrm{d}^3x  \left[\frac{1}{2m}\nabla\bPsi^\dagger\cdot\nabla\bPsi + V_\mathrm{nl}\right]\,,
\end{align}
which is the sum of the kinetic and potential energy, and can be obtained from the nonrelativistic action. The total energy, $E= m N+\mathcal{E}$, includes the rest mass energy and is the appropriate approximation to equation~\eqref{eq:E_relativistic}. Here, $N \equiv \int\mathrm{d}^3x\,\bPsi^\dagger\bPsi$ is the conserved particle number resulting from the $\bPsi \rightarrow e^{i\alpha}\bPsi$ symmetry of nonrelativistic Lagrangian \eqref{eq:NR_action}.

The equation of motion for $\bPsi$ is a nonlinear Schr\"{o}dinger equation $i\partial_t\bPsi=-\nabla^2\bPsi/(2m)+\partial_{\bPsi^\dagger} V_\rm{nl}$.

\section{Oscillon solutions}
The ground state solution of this nonlinear Schr\"{o}dinger equation with a fixed particle number $N=N_\star$ can be obtained by extremizing $\mathcal{E}+\mu (N-N_\star)$ where $\mu$ is a Lagrange multiplier. Such a solution must have the form
\begin{align}
    \bPsi(t,\bx)=\bm{\Psi}(\bx) e^{i\mu t},
\end{align}
where the profile $\bm{\Psi}$ satisfies
\begin{align}
\label{eq:profilebPsi}
    -\mu \bm{\Psi}=-\frac{1}{2m}\nabla^2\bm{\Psi}+\partial_{\bm{\Psi}^\dagger} V_\rm{nl}.
\end{align}
Note that in a Cartesian basis, $\bm{\Psi}(\bx)=\sum_{j=1}^3\psi_j(\bx)e^{i\phi_j(\bx)}\hat{\bx}_j$, and $\psi_j$ and $\phi_j$ are real valued functions. 
The profile equation \eqref{eq:profilebPsi} contains a set of 6 equations for these 6 real functions.

We now hunt for the lowest energy, spatially localized solutions for a fixed particle number, keeping in mind that there might be multiple solutions that are local minima of the energy. We do not know \textit{a priori} which one is the true ground state.

The spatial variation in the phases $\phi_j(\bx)$ costs gradient energy, so we will set these to be spatially independent. Thereafter, by shifting the time coordinate, we can always set one of these three phases (say $\phi_z$) to zero. We are then left with the task of determining two phases $\phi_{x,y}$ and three spatially varying functions $\psi_{x,y,z}$.

Like the phase, the spatial variation of the direction of the vector field also costs gradient energy. As a result, we consider vector field configurations that point in the same direction at a given instant of time. We will restrict our attention to configurations with a spherically symmetric energy density. With these considerations, we focus on the following form of the field configuration: 
\begin{equation}
\label{eq:Cartesian}
\bm{\Psi}(\bx)=\psi_x(r)e^{i\phi_x}\hat{\bx}+\psi_y(r)e^{i\phi_y}\hat{\by}+\psi_z(r)\hat{\bz},
\end{equation} 
where $r=\sqrt{x^2+y^2+z^2}$.

\begin{figure}
	\centering
	\includegraphics[width=0.9
	\linewidth]{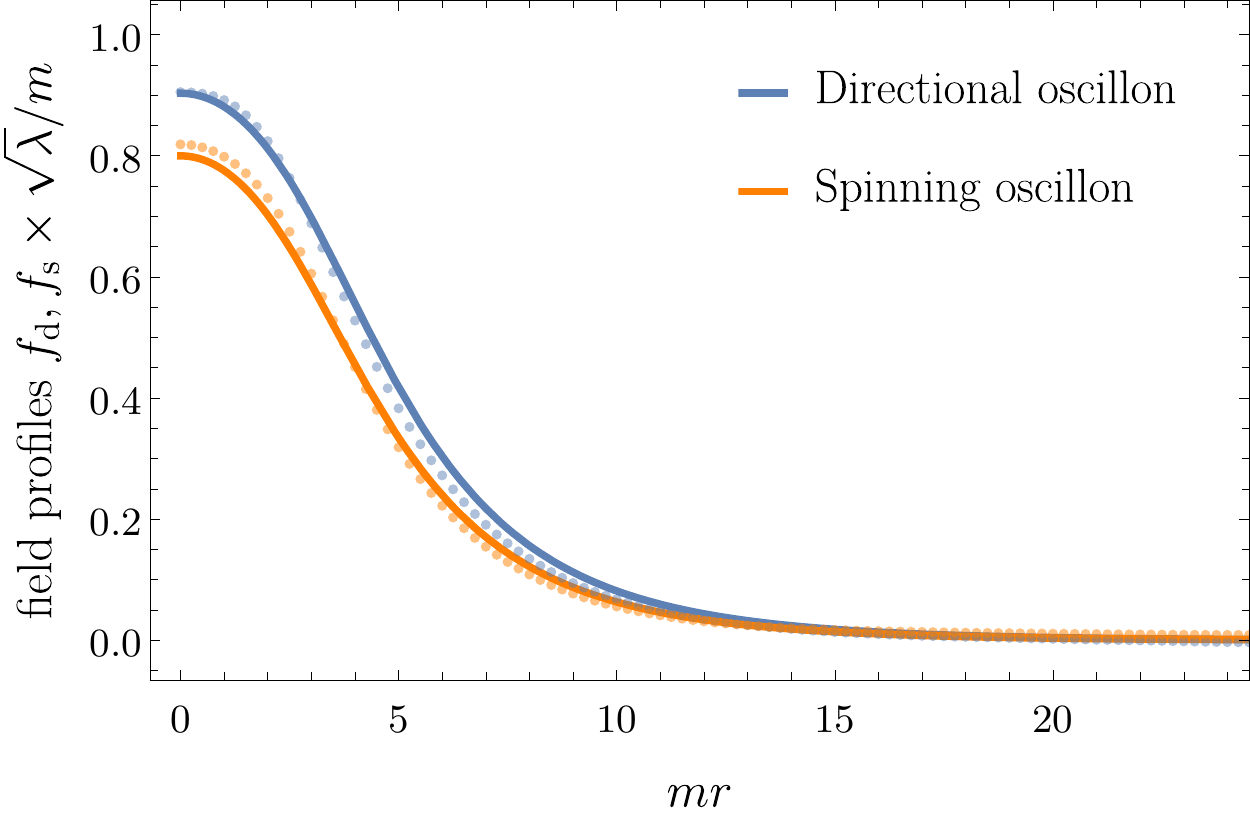}
	\caption{Solid lines show the spatial profiles for directional and spinning oscillons derived using the nonrelativistic theory. Dots represent the appropriately averaged profiles extracted from simulations.  For these profiles, we have $\omega=m-\mu\approx 0.975\,m$.}
	\label{fig:NRprofiles_ext_pol}
\end{figure}

Upon substituting equation \eqref{eq:Cartesian} into equation \eqref{eq:profilebPsi}, we get strong restrictions on the phases and profiles. Specifically, only two distinct classes of oscillons are allowed, which we refer to as directional and spinning oscillons respectively. These classes are represented by
\begin{align}
&\sqrt{\frac{2}{m}}\bm{\Psi}_{\rm{d}}(\bx)=f_{\rm{d}}(r)\hat{\bz}\,,\\
&\sqrt{\frac{2}{m}}\bm{\Psi}_{\rm{s}}(\bx)=f_{\rm{s}}(r)(\hat{\bx}+i\hat{\by}),
\end{align} where $f_{\rm{d}}$ and $f_{\rm{s}}$ satisfy the following profile equations
\begin{align}\label{eq:profile_d}
    -\mu f_{\rm{d}}=&-\frac{1}{2m}\nabla^2f_{\rm{d}}-\frac{3\lambda}{8m}f_{\rm{d}}^3+ \frac{5\gamma}{16m}f_{\rm{d}}^5\,,\\
        -\mu f_{\rm{s}}=&-\frac{1}{2m}\nabla^2f_{\rm{s}}-\frac{\lambda}{2m}f_{\rm{s}}^3+ \frac{\gamma}{2m}f_{\rm{s}}^5\,.
\end{align}
All other oscillons (with radially symmetric field components) in a given class are spatial rotations of these representative oscillons.  The nodeless and spatially localized solutions can be obtained by a numerical shooting method and are shown in Fig.~\ref{fig:NRprofiles_ext_pol}.

%~~~~~~~~~~~~~~~~~~
\begin{figure*}
	\centering
	\includegraphics[width=1\linewidth]{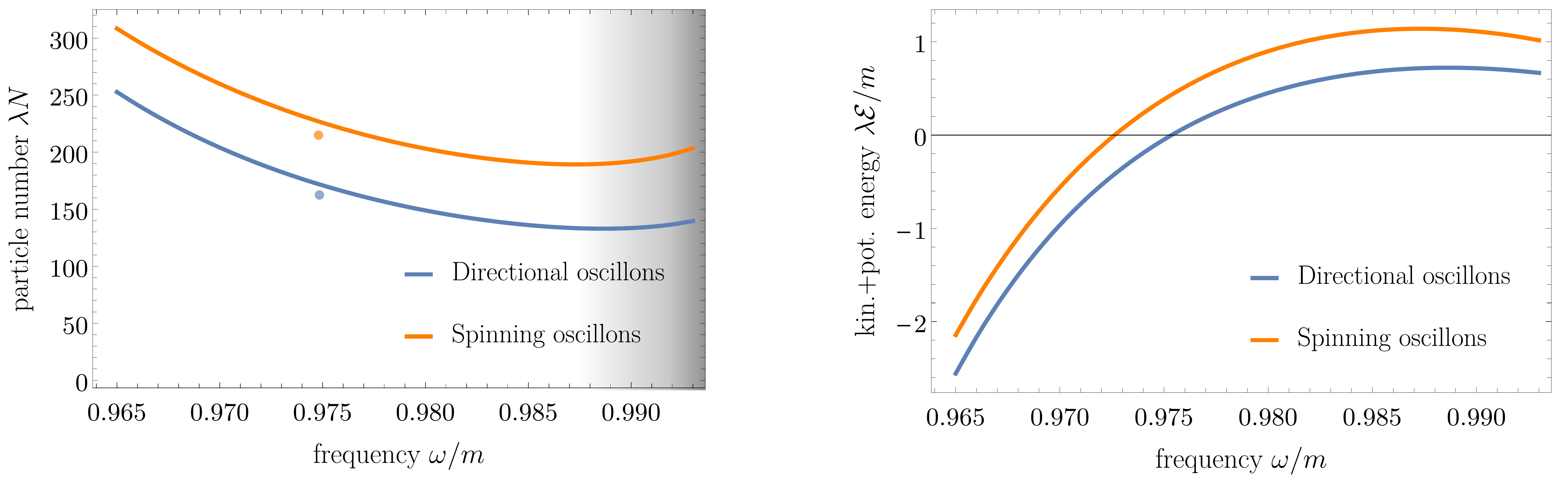}
	\caption{Left panel: Total particle number $N$ vs frequency $\omega=m-\mu$ for the two oscillons. The minimum of each curve determines the respective $\omega$ values below which the oscillons start to exhibit classical stability. The dots indicate the values obtained for oscillons from lattice simulations. Right panel: The sum of the kinetic and potential energy $\mathcal{E}$ as a function of $\omega$. Negative $\mathcal{E}$ represents bound objects and hence ``quantum stability'' in the nonrelativistic limit.}
	\label{fig:E_vs_omega_ext_pol}
\end{figure*}
%~~~~~~~~~~~~~~~~~~

These two oscillons are also readily apparent if we  decompose the field $\bm{\Psi}$ in an orthonormal polarization basis with respect to a fixed direction (see \cite{Jain:2021pnk} for an explicit discussion). That is, $\bm{\Psi}(\bm{x})=\sum_{m_s}\psi^{(m_s)}(\bx)\bepsilon^{(m_s)}_{\hat{z}},$ where $m_s={-1,0,1}$ are the spin multiplicities, and  $\bepsilon^{(0)}_{\hat{z}}=\hat{\bz},\bepsilon^{(\pm 1)}_{\hat{z}}=(\hat{\bx}\pm i\hat{\by})/\sqrt{2}$. Then, the directional and spinning oscillons are the ones with $\bm{\Psi}_{\rm{d}}(\bx)\propto\bepsilon^{(0)}_{\hat{z}}$ and $\bm{\Psi}_{\rm{s}}(\bx)\propto\bepsilon^{(+1)}_{\hat{z}}$, respectively \footnote{In general for an arbitrary direction $\hat{n}$, the set $\{\bepsilon_{\hat{n}}^{(m_s)}\}$ is orthonormal in the sense that ${\bepsilon}^{(m_s)\;\dagger}_{\hat{n}}\cdot{\bepsilon}^{(m_s')}_{\hat{n}} = \delta_{m_s,m_s'}$, and $\sum_{m_s}\left[{\bepsilon}^{(m_s)}_{\hat{n}}\;{\bepsilon}^{(m_s)\dagger}_{\hat{n}}\right]_{ij} = \delta_{ij}$. Also note that the $m_s=-1$ state can be obtained by a rotation of the $m_s=1$ state.}.

While the spin density for the directional oscillon is zero, for the spinning oscillon it is given by $\bS_{\rm{s}}=i\bPsi_{\rm{s}}\times \bPsi_{\rm{s}}^\dagger=mf_{\rm{s}}^2(r)\hat{\bz}$.  The total spin of the configurations are given by
\begin{equation}
    \bS_{\rm{d}}^{\rm{tot}}=0\,,\quad \bS_{\rm{s}}^{\rm{tot}}=N\hat{\bz}\,,
\end{equation}
where $N$ is the particle number. Note that $N=\mathcal{O}[10^2]/\lambda$ can be macroscopically large for $\lambda\ll 1$ (see Fig.~\ref{fig:E_vs_omega_ext_pol}). The orbital angular momentum vanishes for both configurations.

In the nonrelativistic limit, the expressions for the real-valued vector field for directional and spinning oscillons are
\begin{align}
    \label{ansatz_directional}
    \bW_{\rm d}(t,\bx)&=f_{\rm d}(r)\cos(\omega t)\hat{\bz}\,,\\
    \label{ansatz_spinning}
    \bW_{\rm s}(t,\bx)&=f_{\rm s}(r)\left[\cos (\omega t) \hat{\bx}+\sin (\omega t)\hat{\by}\right]\,,
\end{align}
where $\omega=m-\mu$. In appendix \ref{app:smallamp} we also provide an ``$\epsilon$-expansion'' scheme to obtain small-amplitude oscillons. Within this expansion, we provide subleading corrections which show deviations from spherical symmetry of the profiles as well as small corrections to the vector directions. However, this scheme (unlike the nonrelativistic expansion in this section) makes it difficult to obtain solutions in the stable regime.

\section{Energy and Stability}
The particle number as well as the energy for these solutions as a function of $\omega=m-\mu$ are shown in Fig.~\ref{fig:E_vs_omega_ext_pol}. From the figures, it is clear that the directional and spinning solutions have different energy for fixed $\omega$. Furthermore, for a fixed $N=N_{\rm{d}}=N_{\rm{s}}$, we have \footnote{For $\gamma=0$, one can show analytically by appropriate scaling of the field and spacetime coordinates that $\mathcal{E}_{\rm{d}}=(4/9)\mathcal{E}_{\rm{s}}$ for $N_{\rm{s}}=N_{\rm {d}}$ with $\mathcal{E}_{\rm{d}},\mathcal{E}_{\rm{s}}>0$. Alternatively for fixed $\omega$, $\mathcal{E}_{\rm{d}}=(2/3)\mathcal{E}_{\rm{s}}$, consistent with what we see at the right edge of the right panel in Fig.~\ref{fig:E_vs_omega_ext_pol}.}
\begin{equation}
\mathcal{E}_{\rm{d}}<\mathcal{E}_{\rm{s}}~.%<\mathcal{E}_{\rm{h}}.
\end{equation}
In contrast, $\mathcal{E}_{\rm{d}}=\mathcal{E}_{\rm{s}}$ for vector solitons supported by gravitational interactions alone \cite{Jain:2021pnk}. The reason for this degeneracy breaking in energy is the $\bm{S}\cdot\bm{S}$ term in $\Vnl$, which is absent in the gravitational case. It also prohibits construction of fractionally polarized solutions via linear superpositions of maximally polarized solitons \cite{Jain:2021pnk}.

As seen in Fig.~\ref{fig:E_vs_omega_ext_pol}, for each solution, there exist regimes where $dN/d\omega <0$, as well as $\mathcal{E}<0$ indicating classical and ``quantum stability'' respectively \cite{Friedberg:1976me, Levkov:2017paj}. Classical stability is the \textit{less restrictive} of the two. Note that this assumes the number changing processes are suppressed as should be the case in the nonrelativistic regime. While suppressed, these processes are present in the relativistic theory and lead to a slow decay of the oscillons via relativistic radiation \cite{Fodor:2008du, Grandclement:2011wz, Mukaida:2016hwd, Ibe:2019vyo, Zhang:2020bec, Zhang:2020ntm, Cyncynates:2021rtf}.

\section{Relativistic Simulations}
Foregoing nonrelativistic approximations, we simulate vector oscillons on a $3+1$-dimensional lattice by discretizing the relativistic equations (see appendix \ref{sec:numerics} for details). We confirm that the directional and spinning oscillons exist in the fully relativistic theory, and are  long-lived compared to their oscillation period.

In order to see that the existence of vector oscillons is not too sensitive to the choice of initial conditions, we use a Gaussian ansatz $F(r)=C e^{-r^2/R^2}$ with $C\lesssim m/\sqrt{\lambda}$ and $R\sim 10\,m^{-1}$ to initialize vector field components for our two different oscillons. Depending on the choice of $C$ and $R$,  the fields latch on to oscillon configurations with different dominant frequency $\omega$ (after an initial transient).  For ease of comparison, we intentionally pick $C$ and $R$ so that in each case we get an oscillon with approximately the same $\omega\approx 0.975\,m$. This frequency is consistent with oscillons being classically stable according to the analysis in the previous section (see Fig.~\ref{fig:NRprofiles_ext_pol}).

For the directional solitons, we start with an initial profile $\bW(t,\bx)|_{t=0}=F(r)\hat{\bz}$ and $\dot{\bW}(t,\bx)|_{t=0}=0$.  Within $t= \mathcal{O}(10^2)\,m^{-1}$, this initial Gaussian profile settles into an oscillon configuration with frequency $\omega \approx 0.975\,m$ and the energy $E_{\rm d}\approx 164\,m/\lambda$.  For this $\omega$, the energy of the oscillon from the nonrelativistic approximation is $E_{\rm d}=mN+\mathcal{E}\approx 171 m/\lambda$ with a radius $R_{1/e}\approx 6m^{-1}$ as seen in Figs.~\ref{fig:E_vs_omega_ext_pol} and ~\ref{fig:NRprofiles_ext_pol} respectively.

As the ansatz \eqref{ansatz_directional} is not fully compatible with the relativistic equations, a small deviation of the field configuration from the $\hat{\bz}$ direction is expected, which is indeed observed in our simulations. See appendix \ref{sec:numerics} for snapshots of numerical profiles. In the quantities we have checked, such as profiles, energy etc., there is typically a few percent fractional difference between the results of the simulations and the nonrelativistic solutions. This difference is consistent with our expectation that relativistic corrections should be of order  $|\nabla^2/m^2|\sim 1/(mR_{1/e})^2=\mathcal{O}(10^{-2})$.

Taking advantage of a cylindrical symmetry exhibited by directinal oscillons, we carry out long-time simulations in effectively $2+1$ dimensions with absorbing boundary conditions. After an initial transient, the oscillon does not show significant energy loss for the duration of the simulations ($\sim 10^5 m^{-1}$). We note that the lifetimes may be longer because of non-trivial suppression in the decay rates as seen in the case of scalar oscillons \cite{Zhang:2020bec, Zhang:2020ntm}.

In order to obtain spinning oscillons, we start the simulation with $\bW(t,\bx)|_{t=0} = F(r)\hat{\bx}$, $\dot{\bW}(t,\bx)|_{t=0} = F(r)\hat{\by}$. With these initial conditions, the field quickly settles into a spinning oscillon configuration with frequency $\omega\approx 0.975\,m$ and the energy $E_{\rm{s}}\approx 216\,m/\lambda$. Our analytic estimates yield $E_{\rm{s}}\approx 225\,m/\lambda$. Along with dominant components in the $x-y$ plane, we see small components in the $\hat{\bz}$ direction. Moreover, the energy density deviates slightly from spherical symmetry. Once again, the analytic estimates from our nonrelativistic theory differ from the results from relativistic simulations by a few percent, consistent with our expectations.

Unlike the directional case, we cannot take advantage of symmetries to do a long-time simulation in effectively lower dimensions. However, we have verified that with absorbing boundary conditions, the spinning oscillon does not decay away for at least $\sim 10^3m^{-1}$.

\section{Discussions}
We have presented two new oscillon solutions in real-valued vector fields with attractive self-interactions. The oscillons are maximally polarized: the directional oscillon has zero intrinsic spin, while the spinning oscillon has maximum intrinsic spin equal to the occupation number of the oscillon in the nonrelativistic limit (i.e. $\b S_{\mathrm{tot}} = \hbar N \hat{n}$). In the case of gravitational interactions alone, the two solitons (in the nonrelativistic limit) are degenerate in energy for fixed particle number, and can be appropriately superposed to form fractionally polarized solitons~\cite{Jain:2021pnk}. Here however, the presence of spin-spin interactions  breaks this degeneracy, making the directional oscillon lower in energy, and furthermore prohibits fractionally polarized solitons.

We have confirmed that these oscillons are not too sensitive to the choice of initial conditions, and furthermore do not decay away for at least $10^3\, m^{-1}$ (see appendix \ref{sec:lifetimes} for further discussion of lifetimes as well as model parameters in two different production mechanisms for the vector field). A more detailed longer-timescale simulation, as well as analytic calculation of the decay rates (similar to \cite{Mukaida:2016hwd, Ibe:2019vyo, Zhang:2020bec, Zhang:2020ntm}) are warranted.
The lack of detailed sensitivity to initial conditions and their long lifetimes make them potentially relevant in astrophysical and cosmological scenarios.

The two oscillon solutions presented in this paper have approximately spherically symmetric energy density but not field configurations. However, there is another oscillon solution for which both the field and energy density are exactly spherically symmetric, known as the hedgehog oscillon~\cite{Adshead:2021kvl,Loginov:2015rya, Brito:2015pxa, Sanchis-Gual:2017bhw, Sanchis-Gual:2019ljs, CalderonBustillo:2020srq}. We find that this solution includes significant relativistic corrections towards its center, and is also higher in energy (and likely harder to form from generic initial conditions) than the two maximally polarized oscillons presented here. This will be the subject of future work.

The spin nature of the vector field, manifest in these oscillons, can lead to novel phenomenological implications. Collisions and mergers of dense vector oscillons can lead to gravitational wave production, which might be distinct from the scalar case ~\cite{Zhou:2013tsa, Palenzuela:2017kcg, Liu:2017hua, Helfer:2018vtq, Amin:2018xfe, Dietrich:2018jov, Lozanov:2019ylm}. If the massive (dark) vector field kinetically mixes with the visible photon, namely $\cal L \supset (\sin\alpha/2) X^{\mu\nu} F_{\mu\nu}$ where $\sin\alpha$ is the mixing parameter and $F_{\mu\nu}$ is the field strength of the photon \cite{Caputo:2021eaa}, collisions between polarized vector oscillons, or interaction with strong magnetic fields can also lead to specific outgoing radiation patterns based on oscillon polarization (see~\cite{Hertzberg:2020dbk,Levkov:2020txo,Amin:2020vja} for scalar case). If such vector oscillons exist today, and interact with terrestrial experiments \cite{Caldwell:2016dcw, Baryakhtar:2018doz, Chiles:2021gxk, Chen:2021bdr}, detectable signatures that depend on the polarization state of the vector field might be possible.

Formation mechanisms and production rates of vector oscillons, along with their early universe implications remain to be explored. The misalignment mechanism for production of dark photon dark matter \cite{Agrawal:2018vin,Co:2018lka}, where an oscillating inflaton or axion field transfers its energy to dark photons efficiently via a resonant instability, could produce vector oscillons resulting in additional small-scale structure in the early universe. Vector oscillons may also form naturally at the end of vector field inflation \cite{Ford:1989me, Golovnev:2008cf, Koivisto:2008xf} analogous to scalar cases \cite{Amin:2011hj,Amin:2019ums}, from ``thermal'' initial conditions \cite{Farhi:2007wj}, or by purely gravitational clustering in the early and contemporary universe \cite{Levkov:2018kau, Schive:2014dra,Blinov:2021axd,Eggemeier:2021smj}.

Beyond their cosmological context, we are currently exploring whether nonrelativisitic vector oscillons with isospin can be realized in multicomponent Bose-Einstein condensates with attractive self-interactions.

\begin{acknowledgments}
We would like to thank Ray Hagimoto, Siyang Ling, Andrew Long and Zong-Gang Mou (Rice University) for useful comments. MJ would like to thank Mark Hertzberg for useful discussions regarding abelian Higgs model. HYZ would also like to thank Lingyuan Ji (Johns Hopkins University) for helpful discussions. This work is supported by a DOE grant DOE-0000250746.
\end{acknowledgments}

\appendix

%%%%%%%%%%%%%%%%%%%%%%%%%%%
\section{A low-energy effective Proca theory}\label{app:Effective_action}
%%%%%%%%%%%%%%%%%%%%%%%%%%%

Owing to the linear growth in energy of the helicity-0 mode of a massive vector degree of freedom, theory of a self-interacting massive vector field $W_{\mu}$ violates perturbative unitarity at high-energy scales \cite{Weinberg:1995mt,Schwartz:2014sze}. In order to restore it in the most conservative way, we need a new particle with mass comparable to the energy scales of unitarity violation, such that it compensates for the aforementioned growth in any scattering process. This can also be recast (and as is usually done) in the language of gauge invariance: At high energies, gauge invariance in $W_{\mu}$ should be restored such that the helicity-0 mode plus this new degree of freedom reorganize themselves into a local $U(1)$ symmetric field structure. This is the symmetry restoration/Higgs mechanism (or the symmetry removal mechanism when climbing down the energy scales). At energies well below the mass of the new particle then, this fetches an effective theory of a massive vector field $W_{\mu}$ with self-interactions. In this appendix, we show that the Abelian-Higgs model indeed produces an attractive potential for $W_\mu$ at energies well below the mass of the Higgs. This is no surprise since a spin-0 particle mediates an attractive force.

Working with the Euler representation of the complex scalar field, and expanding the radial degree of freedom (Higgs) around the vev $v$, the Lagrangian density in Unitary gauge is
\begin{align}
	\nonumber
	\mathcal{L} =& -\frac{1}{4}X_{\mu\nu}X^{\mu\nu} - \frac{1}{2}m^2W_\mu W^\mu - \frac{1}{2}\partial_{\mu}h\,\partial^{\mu}h \\
	\nonumber
	&- \frac{1}{2}M^2h^2 - gmhW_{\mu}W^{\mu} - \frac{1}{2}g^2h^2W_{\mu}W^{\mu} \\
	&- \frac{1}{2}\frac{gM^2}{m}h^3 - \frac{1}{8}\frac{g^2M^2}{m^2} h^4\,.
\end{align}
Here $m$, $M$, and $g$ are the mass of the vector field $W_\mu$, mass of the Higgs field $h$, and the gauge coupling constant respectively. The masses are related to the vev $v$, the Higgs self-coupling $\tilde{\lambda}$, and the gauge coupling $g$ as $M = v\sqrt{2\tilde{\lambda}}$ and $m = gv$. Now, we assume the hierarchy $m \ll M$ (or equivalently $g \ll \sqrt{2\tilde{\lambda}}$), in order to be able to integrate out the Higgs. Note that this mass hierarchy is consistent with the quantum theory: Radiative corrections to the gauge coupling $g$ are proportional to it's bare value. Therefore, we can safely assume it to be small. Whereas on the other hand, even if $M$ were zero, the quartic self-coupling of the Higgs generates a mass term (through the tadpole diagram), pushing $M$ away towards the cut-off of the theory. This can also be seen from the behavior of global symmetries. In the limit $g \rightarrow 0$ while keeping $M = v\sqrt{2\tilde{\lambda}}$ fixed, $W_{\mu}$ decouples with the two scalars and we recover a global U(1) (note that there \textit{isn't} a global U(1) in the Higgs phase~\cite{Hertzberg:2018kyi,Hertzberg:2019ffc}).

To get an effective theory at energies well below $M$, we wish to integrate out the Higgs field. At tree level, we can simply plug back the solution to the Higgs equation of motion (discarding the derivative term $\Box h$) into the Lagrangian density. Upon doing so we get the following effective Lagrangian density for the vector field
\begin{align}
	\mathcal{L}_{\mathrm{eff}} = -\frac{1}{4}X_{\mu\nu}X^{\mu\nu} - \frac{1}{2}m^2W_{\mu}W^{\mu} + \frac{1}{4}\lambda(W_{\mu}W^{\mu})^2 + ...
\end{align}
where $\lambda = 2g^2m^2/M^2$, and the `...' represents derivative couplings (suppressed by factors of $M$).

We would like to thank Mark Hertzberg for pointing to us an error in the previous version of this draft, in that the abelian Higgs model only generates a quartic interaction term upon integrating out the heavy Higgs. In this work, we have assumed a sextic term as well in order to study vector oscillons. We remain agnostic towards a mechanism that generates such higher order (sextic and above) self-interaction terms. Additionally, different coefficients of higher-dimensional couplings may be possible under different symmetry restoration setups in the UV. Besides the Higgs mechanism, self-interactions could also arise in vector Galileons \cite{Heisenberg:2014rta} and in the nonrelativistic limit of vector fields that are non-minimally coupled to gravity.

It is also important to point out that in 3+1 dimensions, the long-lived oscillon solutions we find in our simulations (and also in our nonrelativistic analytic approximation) have field amplitudes in the center that are not too small, meaning $W \sim m/\sqrt{\lambda}$ (see fig.~\ref{fig:NRprofiles_ext_pol}). In this case it may no longer be valid to truncate the potential up to the sextic coupling, and a further analysis is needed to take higher-dimensional couplings into account. We leave this for future work.

Recently it is realized that self-interacting vector fields may yield some singularity problem where the theory breaks down \cite{Mou:2022hqb, Clough:2022ygm}. This implies that vector oscillons are unlikely to have very large amplitudes. We confirm that this problem is not encountered in this work and does not affect our main conclusions.

%%%%%%%%%%%%%%%%%%%%%%%%%%%%%%%%%%%%%%%%%
\section{Small-amplitude expansions}
%%%%%%%%%%%%%%%%%%%%%%%%%%%%%%%%%%%%%%%%%
\label{app:smallamp}

For small-amplitude scalar oscillons, usually an `$\epsilon$ expansion scheme' is employed~\cite{Fodor:2008es, Amin:2010jq, Amin:2013ika}. In this appendix, we present this for our vector oscillons. Introducing a dimensionless $\epsilon$ parameter, we re-scale space, time and the field as 
\begin{align}
	\label{dimensionalization}
	\nonumber
	\bx \rightarrow \frac{\bx}{m\epsilon}\qquad t \rightarrow \frac{t}{m\sqrt{1-\epsilon^2}},\qquad \\
	W_\mu = \frac{m}{\sqrt{\lambda}}\left[\epsilon W_\mu^{(1)} + \epsilon^2 W_\mu^{(2)} + \epsilon^3 W_\mu^{(3)} + \cdots\right].
\end{align}
We also scale $\gamma \rightarrow (\lambda^2/m^2)\gamma$. Note that the new variables are all dimensionless.
By plugging this expansion into the field equation~\eqref{EOM1} and collecting terms up to the order $\cal O(\epsilon^4)$, we obtain
\begin{align}
	\label{EOM_Wi_order1}
	\ddot{\b W}^{(1)}+{\b W}^{(1)}&=0 ~,\\
	\label{EOM_Wi_order2}
	\ddot{\b W}^{(2)}+{\b W}^{(2)}&=0 ~,\\
	\label{EOM_Wi_order3}
	\ddot{\b W}^{(3)}+{\b W}^{(3)}&=\left[\nabla^2-\left(1-{\b W}^{(1)}\cdot{\b W}^{(1)}\right)\right]{\b W}^{(1)} ~,
\end{align}

\begin{widetext}
\begin{align}
	\label{EOM_Wi_order4}
	\ddot{\b W}^{(4)}+{\b W}^{(4)}
	&=\left[\nabla^2-\left(1-{\b W}^{(1)}\cdot{\b W}^{(1)}\right)\right]{\b W}^{(2)}
	+\left[2{\b W}^{(1)}\cdot{\b W}^{(2)}\right]{\b W}^{(1)} ~,\\
	\label{EOM_Wi_order5}
	\nonumber
	\ddot{\b W}^{(5)}+{\b W}^{(5)}
	&=\left[\nabla^2-\left(1-{\b W}^{(1)}\cdot{\b W}^{(1)}\right)\right]{\b W}^{(3)}
	+\left[2{\b W}^{(1)}\cdot{\b W}^{(2)}\right]{\b W}^{(2)}\\
	&+\left[\nabla^2-\left\{1-\left({\b W}^{(1)}\cdot{\b W}^{(1)}-(\nabla\cdot \dot{\b W}^{(1)})^2+{\b W}^{(2)}\cdot{\b W}^{(2)}+2{\b W}^{(1)}\cdot{\b W}^{(3)}\right)\right\}\right]{\b W}^{(1)}\nonumber\\
	&+\gamma({\b W}^{(1)}\cdot{\b W}^{(1)})^2{\b W}^{(1)} -\nabla\left[\left\{\partial_t({\b W}^{(1)}\cdot{\b W}^{(1)})\partial_t\nabla+\nabla({\b W}^{(1)}\cdot{\b W}^{(1)})\right\}\cdot{\b W}^{(1)}\right].
\end{align}
\end{widetext}
Here we have used $W_0^{(n)}=-\nabla\cdot\dot{\b W}^{(n-1)}$ for $n\le 3$ based on~\eqref{EOM1} and $\b W^{(0)}=0$. We find that we can consistently set $\b W^{(n)} = 0$ for $n =$ even. Note that solutions of~\eqref{EOM_Wi_order1} must have the form $\bW^{(1)}=\sum_{j=1}^{3} w_j(\bx)\Re[e^{it+\phi_j(\bx)}]\hat{\bx}_j$. For lowest energy, spatially localized solutions, we set $\phi_j(\bx)=\phi_j$ and $w_j(\bx)=w_j(r)$. Once we plug this expression into \eqref{EOM_Wi_order3}, and make sure that the right hand side does contain any terms proportional to $\cos t$ or $\sin t$, we get severe restrictions of the phases $\phi_j$ and profiles $w_j(r)$. The reason for removing terms proportional to $\sin t$ and $\cos t$ is so that $\bW^{(3)}$ does not grow with time via a forced resonance. There are only two possibilities for $\bW^{(1)}(r,t)$ given our assumptions (up to spatial rotations, or global phase shifts). The first one is $\bW^{(1)}=v_\textrm{d}(r)\sin t \,\hat{\bz}$ which is the directional oscillon, and the second is $\bW^{(1)}=v_\textrm{s}(r)[\cos t \,\hat{\bx}+\sin t \,\hat{\by}]$ which is the spinning oscillon.  The profile equations are provided below.

By repeating this exercise at every order in the $\epsilon$ expansion, we can find profiles to arbitrary order. In this appendix we provide the small-amplitude profiles up to sub-leading order for both of our oscillons. The subleading profiles reveal the departures from spherical symmetry, which we also see in our $3+1$ dimensional numerical simulations (see Fig.~\ref{fig:largeprofiles}).

For the directional oscillon,
\begin{align}
	\label{EOM_vd}
	\nonumber
	\b W^{(1)}(r,t) = v_\rm{d}(r) \sin t\,\hat{\b z}\,\quad\rm{where}\\
	\quad\pd_{r}^2 v_\rm{d} + \frac{2}{r}\pd_{r} v_\rm{d} - v_\rm{d} + \frac{3}{4} v_\rm{d}^3 = 0 ~.
\end{align}
The profile equation can be  solved numerically for node-free, localized solutions. The result in shown in Fig.~\ref{fig:subleadingprofile}. To obtain subleading $\bW^{(3)}$ we make use of~\eqref{EOM_Wi_order5} and obtain profile equations by setting the coefficients of terms proportional to $\cos t$ and $\sin t$ on the right hand side equal to zero. After some work, we find that the only possibility is
\begin{widetext}
\begin{align}
	\label{W3_directional_ansatz}
	\b W^{(3)}({\bf x},t) = \frac{\rho z}{{r}^2} f_1\,\sin t\,\hat{\b \rho} + \left[\left(\frac{{z}^2}{{3r}^2} (f_2+2f_3) + \frac{{\rho}^2}{{3r}^2} (f_2-f_3)\right)\sin t - \frac{1}{32}v_{\textrm{d}}^3\sin 3t\right]\hat{\b z}\,,
\end{align}
where $f_n$ (assumed to be spherically symmetric) obey
the following set of PDEs
%~~~~~~~~~~~~~~~~~~~~~~~~~~~~~~
\begin{align}
	\( \nabla^2 - 1 - \frac{6}{{r}^2} + \frac{3}{4} v_\rm{d}^2 \) f_1 &=\[ 4(\pd_{r}v_\rm{d})^2 + 2v_\rm{d}\pd_{r}^2 v_\rm{d} -\frac{2}{r} v_\rm{d}\pd_{r} v_\rm{d} \]v_\rm{d}  ~,\\
	\( \nabla^2 - 1 +  \frac{9}{4} v_\rm{d}^2 \) f_2 &=\[ \frac{9}{128}+ \frac{15}{8}\gamma \]v_\rm{d}^5 +\[ \frac{17}{4} (\pd_{r}v_\rm{d})^2 + 2v_\rm{d} \pd_{r}^2 v_\rm{d} + \frac{4}{r} v_\rm{d} \pd_{r}v_\rm{d} \] v_\rm{d} ~,\\
	\( \nabla^2 - 1 - \frac{6}{r^2} + \frac{9}{4} v_\rm{d}^2 \) f_3 &= \[ \frac{17}{4} (\pd_{r}v_\rm{d})^2 + 2v_\rm{d} \pd_{r}^2 v_\rm{d} - \frac{2}{r} v_\rm{d} \pd_{r}v_\rm{d} \] v_\rm{d} ~,
\end{align}
with $\nabla^2 = \pd_{r}^2 + (2/r)\pd_{r}$. The localized solutions of these equations for $\gamma=1$ are shown in Fig.~\ref{fig:subleadingprofile}.

For the spinning oscillon,
\begin{align}
	\label{EOM_vs}
	\bW^{(1)} = v_\rm{s}(r) (\cos t\,\hat{\b x} + \sin t\,\hat{\b y})\quad\rm{where}\quad
	\pd_{r}^2 v_\rm{s} + \frac{2}{r}\pd_{r} v_\rm{s} - v_\rm{s} + v_\rm{s}^3 &= 0 ~.
\end{align}
The numerically obtained profile is shown in Fig.~\ref{fig:subleadingprofile}. To obtain the sub-leading profile, we once again need to make sure that in~\eqref{EOM_Wi_order5}, there are no source terms with unit frequency. After lengthy algebra, we find that the only possible ansatz for $\b W^{(3)}$ is
\begin{align}
	\label{W3_spinning_ansatz}
	\b W^{(3)} &= \frac{1}{2}\left[\left(\frac{(x^2-y^2)}{r^2}g_1 + \frac{1}{3}g_2 + \frac{2z^2-x^2-y^2}{3r^2}g_3\right) \cos t + \left(\frac{xy}{r^2}g_4\right) \sin t\right]\hat{\b x}\nonumber\\
	&+ \frac{1}{2}\left[\left(\frac{xy}{r^2}g_4\right)\cos t + \left(-\frac{(x^2-y^2)}{r^2}g_1 + \frac{1}{3}g_2 + \frac{2z^2-x^2-y^2}{3r^2}g_3\right) \sin t\right]\hat{\b y}\nonumber\\
	&+ \left[\left(\frac{xz}{r^2}g_5\right) \cos t + \left(\frac{yz}{r^2}g_5\right) \sin t\right]\hat{\b z},
\end{align}
where the $g_n$ (assumed to be spherically symmetric) obey the following set of PDEs:
\begin{figure}
	\centering
	\includegraphics[width=0.47\linewidth]{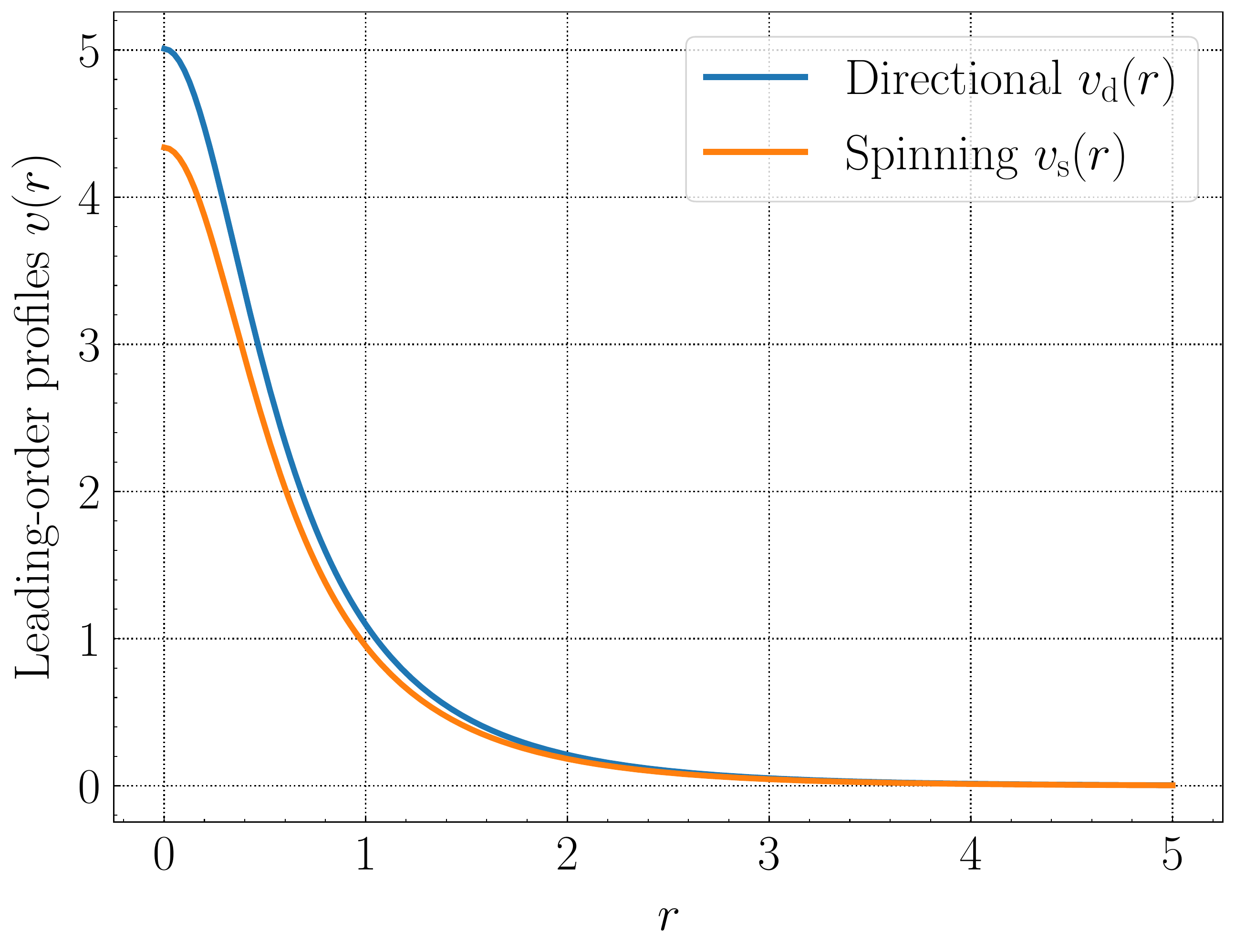}
	\includegraphics[width=0.49\linewidth]{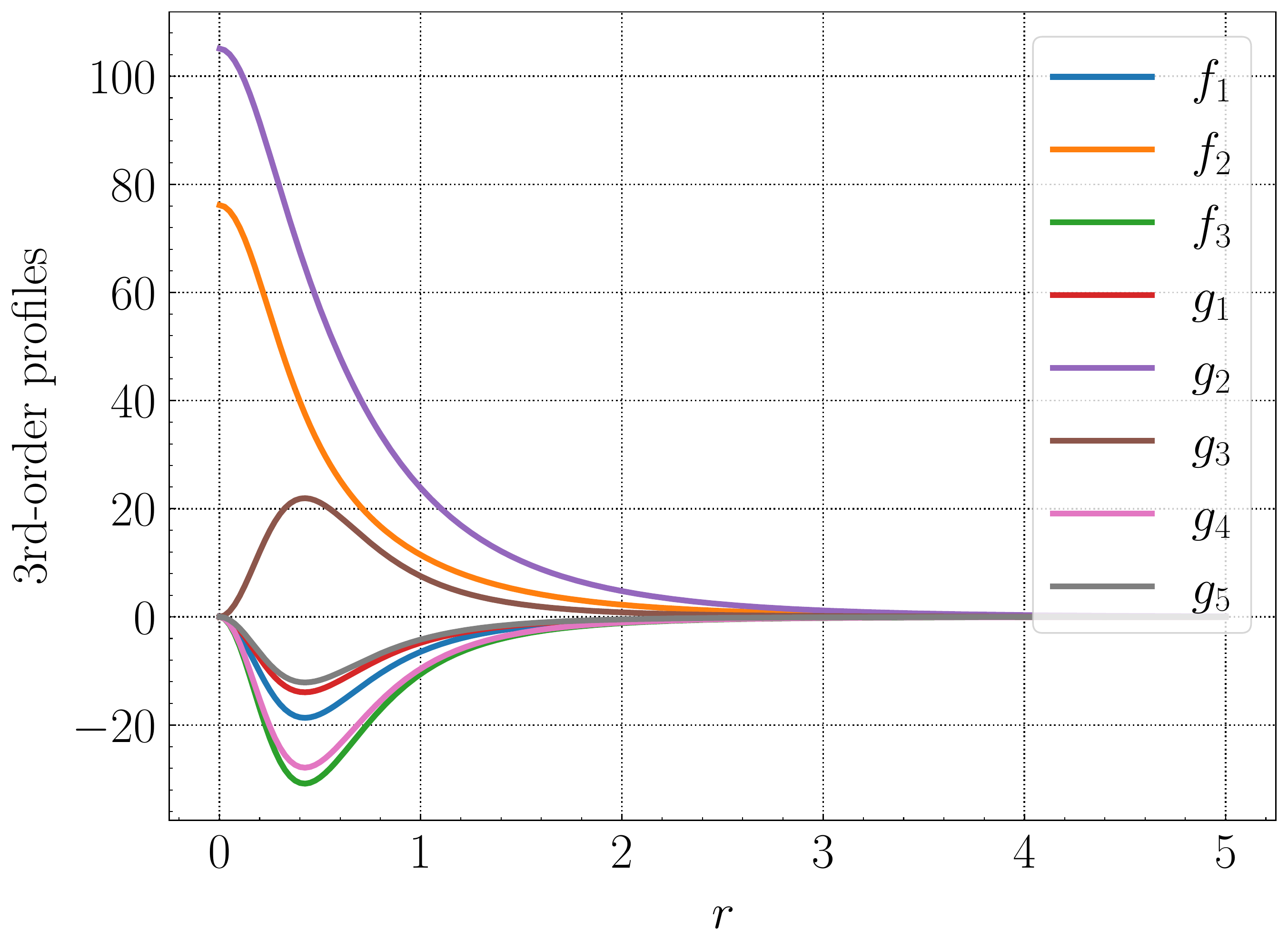}
	\caption{Left: Leading-order profiles, defined in equations \eqref{EOM_vd} and \eqref{EOM_vs}, for directional and spinning oscillons with small amplitudes. Right: Functions $f_n$ and $g_n$ that appear in their sub-leading profiles. As defined in equations \eqref{W3_directional_ansatz} and \eqref{W3_spinning_ansatz}, $f_n$ and $g_n$ characterize the deviation of the spatial profiles from radial symmetry. All quantities shown in the figures are dimensionless. For any $\epsilon \ll 1$, to recover the original dimensional vector profiles, $v_\mathrm{d}$ and $v_\mathrm{s}$ must be multiplied by $\epsilon m/\sqrt{\lambda}$, while the functions $f_n$ and $g_n$ should be multiplied by $\epsilon^3 m/\sqrt{\lambda}$. Similarly the x-axis should be multiplied by $(m\epsilon)^{-1}$ to get the physical spatial extent. These are small-amplitude (compared to $m/\sqrt{\lambda}$) and spatially broad (compared to $m^{-1}$) profiles.}
	\label{fig:subleadingprofile}
\end{figure}
%~~~~~~~~~~~~~~~~~~~~~~~~
\begin{align}
	\( \nabla^2 - 1 - \frac{6}{{r}^2} + 2 v_\rm{s}^2 \) g_1 - 2 v_\rm{s}^2 \( \pd_{r}^2 - \frac{2}{r}\pd_{r} \) v_\rm{s} = 0 ~,\\
	\( \nabla^2 - 1 + 3 v_\rm{s}^2 \) g_2 - \[ 10 (\pd_{r} v_\rm{s})^2 + 4 v_\rm{s} \pd_{r}^2 v_\rm{s} + \frac{8}{r} v_\rm{s}\pd_{r} v_\rm{s} \] v_\rm{s} - 6 \gamma v_\rm{s}^5 = 0 ~,\\
	\( \nabla^2 - 1 - \frac{6}{{r}^2} + 3 v_\rm{s}^2 \) g_3 +  \[ 5 (\pd_{r} v_\rm{s})^2 + 2 v_\rm{s} \pd_{r}^2 v_\rm{s} - \frac{2}{r} v_\rm{s}\pd_{r} v_\rm{s}\] v_\rm{s} = 0 ~,\\
	\( \nabla^2 - 1 - \frac{6}{{r}^2} + 2 v_\rm{s}^2 \) g_4 -  \[ 7 (\pd_{r} v_\rm{s})^2 + 4 v_\rm{s} \pd_{r}^2 v_\rm{s} -\frac{4}{r} v_\rm{s}\pd_{r} v_\rm{s} \] v_\rm{s} = 0 ~,\\
	\( \nabla^2 - 1 - \frac{6}{{r}^2} +  v_\rm{s}^2 \) g_5 - \[ 4 (\pd_{r} v_\rm{s})^2 + 2 v_\rm{s} \pd_{r}^2 v_\rm{s} -\frac{2}{r} v_\rm{s}\pd_{r} v_\rm{s}\] v_\rm{s} = 0 ~,
\end{align}
\end{widetext}
and $\nabla^2 = \pd_{r}^2 + (2/r)\pd_{r}$. The localized solutions of these equations for $\gamma = 1$ are shown in Fig.~\ref{fig:subleadingprofile}.

\twocolumngrid
%%%%%%%%%%%%%%%%%%%%%%%%%%%%%%%%%%
\section{Numerical algorithms}
\label{sec:numerics}
%%%%%%%%%%%%%%%%%%%%%%%%%%%%%%%%%%

In this appendix we describe our numerical algorithms in some detail. We begin with $3+1$-dimensional simulations, which we use for shorter time-scale simulations for both types of oscillons. We then provide details of the simplified simulation for directional oscillons by explicitly imposing cylindrical symmetry.

\begin{figure*}
	\centering
	\includegraphics[width=0.9\linewidth]{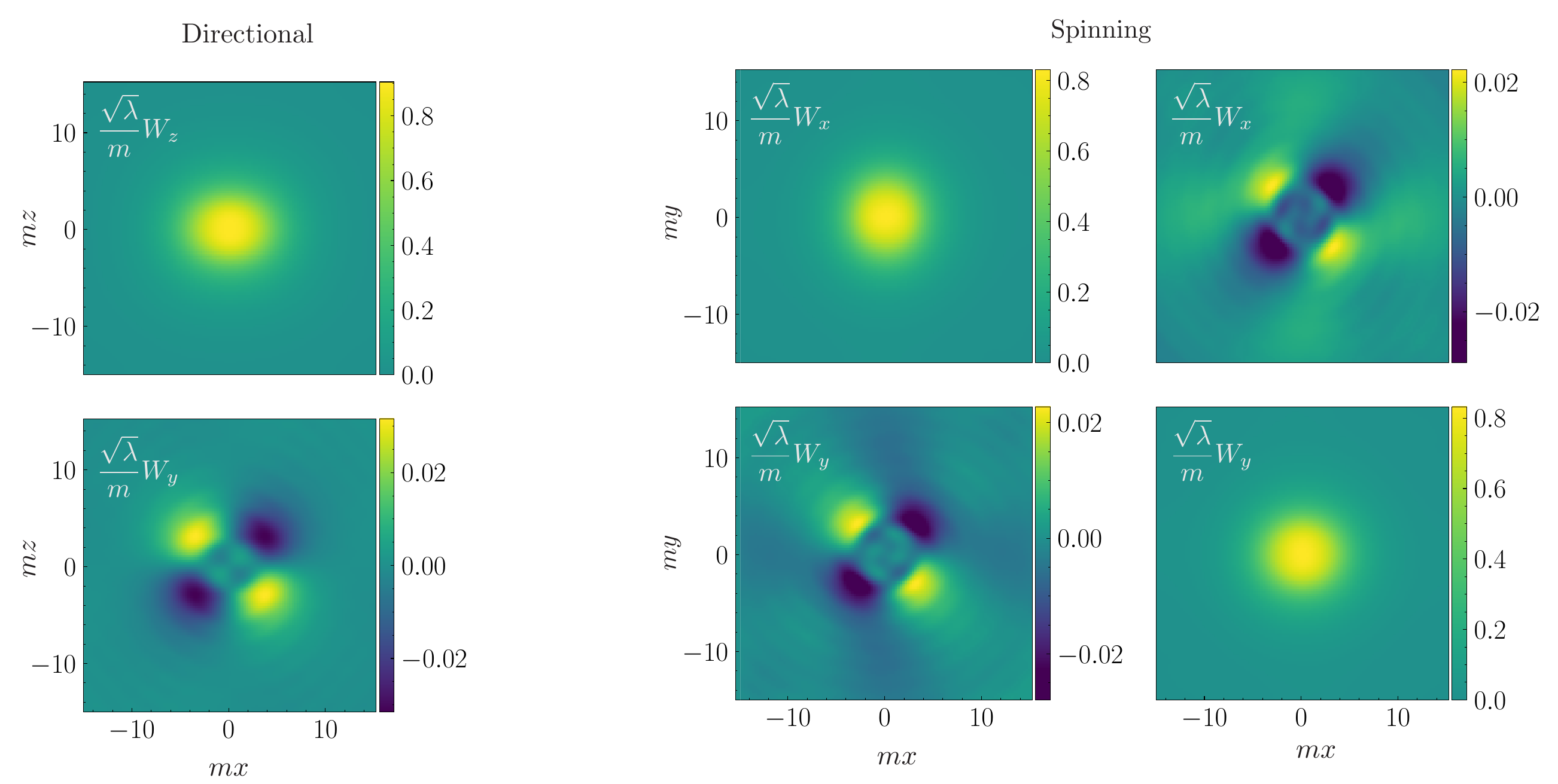}
	\caption{Left: The top panel is a snapshot of the profile of the $z$ component of a directional oscillon (with $\bW$ pointing predominantly in the $\hat{\bz}$ direction). The time is chosen so that the field component has a maximal central value. Bottom panel is the profile for the $y$ component of the field at this same time. Note that these profiles are provided on the $y=0$ plane. On the $z=0$ plane, the $x,y$ components vanish. Right: Snapshot of spatial profiles of the $x$(top) and $y$(bottom) components of $\bW$ for spinning oscillons, with $\bW$ rotating predominantly in the $x$-$y$ plane. In the first column, the time is chosen so that the $W_x$ is at its maximum in the center, whereas for the second column $W_y$ is at its maximum. Note that the deviation from spherical symmetry of the profile of dominant component is small. The subdominant component is small, and does not have spherically symmetric profiles. These results are consistent with both the nonrelativistic solutions used in the main body of the paper. They are also qualitatively consistent with the small-amplitude expansions in appendix \ref{app:smallamp}, in particular, the relative amplitude and shape of the subdominant components.}
	\label{fig:largeprofiles}
\end{figure*}

We first note that besides the Euler-Lagrange equation, one more useful equation can be obtained by noting that $X^{\mu\nu}$ is antisymmetric, i.e.
\begin{align}
	\label{EOM3}
	\pd_\nu \[ 2V'(W_\mu W^\mu) W^\nu \] = 0 ~.
\end{align}
This, along with the Euler-Lagrange equation $\partial_{\mu} X^{\mu\nu} =2V'(W_{\mu}W^{\mu})W^{\nu}$, can be re-written for each component of $W_\mu$ and its time derivative $U_\mu\equiv \dot W_\mu$, i.e.
\begin{align}
	\label{numerical_W0}
	\pd_0 W_\mu &= U_\mu ~,\\
	\pd_0 U_i &= \pd_i U_0 - \pd_i \pd_j W_j + \pd_j\pd_j W_i - 2 V' W_i ~,\\
	\label{numerical_U0}
	U_0 &= \frac{\pd_i(2V' W_i) - 4V'' W_0(W_i U_i)}{2V'-4V'' W_0^2} ~,
\end{align}
where we get one algebraic equation and seven PDEs that involve only 1st-order time derivatives. In addition, we may use the 0-component Euler-Lagrange equation as a  constraint to check numerical accuracy, i.e.
\begin{align}
	\label{numerical_delta}
	\delta \equiv -\pd_i\pd_i W_0 + \pd_i U_i + 2V' W_0 = 0 ~.
\end{align}
Among each component of $W_\mu$ and $U_\mu$, six functions needs to be specified initially and the other two can be solved for by using \eqref{numerical_U0} and \eqref{numerical_delta}. The set of equations can then be numerically evolved by using centered difference in space and a modified iterative Crank-Nicholson method for time, i.e. the type of equation $\pd_t f = S(f, \pd_i f)$ is integrated in the following way \cite{Alcubierre:2003sx}:
\begin{align}
	f^{(1)} &= f^m + \frac{dt}{3} S(f^m, \pd_i f^m) ~,\\
	f^{(2)} &= f^m + \frac{dt}{2} S(f^{(1)}, \pd_i f^{(1)}) ~,\\
	f^{m+1} &= f^m + dt ~S(f^{(2)}, \pd_i f^{(2)}) ~,
\end{align}
where $m$ is the time index and $f^{(n)}$ denotes the $n$-th intermediate variables. This is a $2$nd-order method. To improve numerical stability, we add 4th-order Kreiss-Oliger dissipation terms with a strength parameter $\epsilon_\mathrm{KO}$ at each time step \cite{alcubierre2008introduction}. As for the boundary, we use periodic boundary conditions to obtain profiles for both oscillons (see figures \ref{fig:NRprofiles_ext_pol} and \ref{fig:largeprofiles}), and implement absorbing boundary conditions (i.e. sponge-layer method \cite{Gleiser:1999tj}) to estimate the lifetime of spinning oscillons. Specifically for the latter, a damping term is added to the dynamical equation if the distance to the center $r$ exceeds some damping radius $r_d$ that is far away from the center:
\begin{align}
	\pd_t f = S(f, \pd_i f) + \epsilon_d \theta(r-r_d) (r-r_d)^2 \nabla^2 f ~,
\end{align}
where $\epsilon_d$ is a small parameter that ensures a slow increase of the damping strength.

Time is discretized with $dt=0.08 m^{-1}$, and we perform the simulation in a periodic box of size $70^3 m^{-3}$ with $281^3$ points (with spatial resolution $dx_i= 0.25 m^{-1}$). Note that the dominant radiation modes have a wavelength $\lambda^{\rm{rad}}_j=2\pi/\sqrt{(j\omega)^2-m^2} \sim \rm{few}\times m^{-1}$ \cite{Zhang:2020bec, Zhang:2020ntm}. We have checked that an $\mathcal{O}(1)$ change of parameters (e.g. $dt$, $dx_i$, $\epsilon_\mathrm{KO}$) does not affect the final profiles significantly. With this setup, we are able to simulate the evolution of oscillons for $\mathcal{O}[10^3]m^{-1}$. Longer time simulations will be possible by parallelizing the algorithm, which will be pursued in the future.
%%%%%%%%%%%%%
%%%%%%%%%%%%

In the directional oscillon case, we can simulate for significantly longer times by taking advantage of the cylindrical symmetry of the oscillon field configurations. Effectively, this reduces our problem to $2+1$ dimensions.

For directional oscillons, it is convenient to work with cylindrical coordinates such that $W_0\equiv W_t$ and $\b W\equiv W_\rho \hat{\b\rho} + W_z \hat{\b z}$, and rewrite numerical equations \eqref{numerical_W0}-\eqref{numerical_delta} for $W_t, W_\rho, W_z$ and their time derivatives $U_t, U_\rho, U_z$. We then discretize the spatial grid by $\rho_n=nd\rho$ and $z_n=n dz$ where $n=0,1,2,\cdots, N$ with $N=500$, and set $dt=d\rho/4=dz/4=0.025\,m^{-1}$. Spatial derivatives are approximated by the 2nd-order centered difference everywhere except at boundaries where a special treatment is needed for obtaining $U_t, U_\rho, U_z$. At the inner boundary, $U_t|_{z=0}= U_\rho|_{\rho=0} = U_\rho|_{z=0} = 0$ since they are odd functions of either $\rho$ or $z$, and we obtain $U_t|_{\rho=0}$ and $U_z|_{\rho=0}$ through the Lagrange interpolation with four data points. For better stability performance, values of $U_z|_{z=0}$ are calculated by discretizing the constraint \eqref{numerical_delta} with a finite difference method. At the outer boundary, $U_\rho$ and $U_z$ are assumed to behave like spherical waves so that the following radiative boundary condition is implemented \cite{Alcubierre:2000xu}
\begin{align}
	\frac{x^j}{r} \pd_t U_\rho + \pd_j U_\rho + \frac{x^j}{r^2} U_\rho = 0 ~,
\end{align}
where $x^j = \rho$ or $z$. In reality, the outgoing radiation is not necessarily spherically symmetric. The above condition, however, is sufficient if the boundary is far enough from the origin so that the angular derivative is much smaller than the radial one \cite{alcubierre2008introduction}. The equation for $U_t$ becomes
\begin{align}
	U_t = \pd_{\rho} W_\rho + \frac{1}{\rho}W_\rho + \pd_z W_z ~.
\end{align}
These equations are discretized by 2nd-order finite difference methods.

We find that the lifetime of directional oscillons is $>10^5m^{-1}$. We have checked that an $\mathcal{O}(1)$ change of parameters (e.g. $dt$, $d\rho$, $dz$, $N$, $\epsilon_\mathrm{KO}$) and the number of interpolation points does not affect final results significantly. However, there are still some limitations. For instance, the Gibbs phenomenon is observed for initial profiles with large amplitudes or widths, which limits our ability to probe vector oscillons with large amplitudes (smaller frequencies). A possible solution could be implementing higher-order methods, or using an adaptive grid to evolve the system instead of a fixed one.

%%%%%%%%%%%%%%%%%%%%%%%%%%%%%%%%%%%%%%%%%
\section{Lifetimes of oscillons}
\label{sec:lifetimes}
%%%%%%%%%%%%%%%%%%%%%%%%%%%%%%%%%%%%%%%%%

In general, lifetimes of oscillons go as $\tau \sim \alpha\,m^{-1} \sim (\alpha/10^5) \left(10^{-21}\rm{eV}/m\right)10^{3} ~\rm{yrs}$ where $\alpha$ can be extremely large depending upon the oscillon configuration and the exact shape of the potential (see~\cite{Cyncynates:2021rtf} for the case of scalar oscillons where the authors find it to be as large as $\sim 10^{18}$). In our simulations (with the choice $\gamma = \lambda^2/m^2$), we find $\alpha_{\rm{d}} \gtrsim 10^5$ for directional oscillons, while $\alpha_{\rm{s}} \gtrsim 10^3$ for classically stable spinning oscillons, at the very least. We note that $\alpha_{\rm{s}}$ is expected to be even larger for oscillons that are ``quantum mechanically stable'' ($\mathcal{E}_{\rm{s}} < 0$ in the non-relativistic limit).

Here we provide rough estimates for the lifetimes of the dark photon oscillons in two different possible scenarios: (i) A misaligned scalar that topologically couples to a dark spin-1 field, starts to oscillate around its minimum when the Hubble drops below its own mass, and dumps its energy into the latter owing to a tachyonic instability~\cite{Co:2018lka, Bastero-Gil:2018uel, Agrawal:2018vin}; (ii) Gravitational particle production of spin-1 particles during/towards the end of inflation~\cite{Graham:2015rva,Kolb:2020fwh,Ahmed:2020fhc}. In both cases, $H \sim m$ at the time of dark photon production, dictating that the lifetimes of oscillons are huge as compared to the age of the Universe {\it at that time}: 
\begin{align}
	\tau \sim \alpha\,m^{-1} \sim \alpha\,H^{-1}\quad \rm{with}\quad\alpha \gg 1\,.
\end{align}
Therefore, and specifically owing to the polarization of these vector oscillons, they could play an important phenomenological role. 

Assuming that the spin-1 field constitutes the whole of dark matter, there is an absolute bound on its mass, $m \gtrsim 10^{-21}$ eV, from Lyman-$\alpha$ observations. In scenario (i) there is no additional bound on $m$, while an additional lower bound exists in scenario (ii), $m \gtrsim 10^{-5}$ eV, to evade overproduction of curvature perturbations. An interesting feature of these scenarios is that in the former it is the transverse mode that is dominantly populated, while in the latter it is the longitudinal mode. We leave the detailed analysis to a future study.

\bibliographystyle{apsrev4-2}
\bibliography{reference}
\end{document}